\documentclass{ieeeaccess}
\usepackage{cite}
\usepackage{amsmath,amssymb,amsfonts}
\usepackage{graphicx}
\usepackage{textcomp}
\usepackage{multirow}
\def\BibTeX{{\rm B\kern-.05em{\sc i\kern-.025em b}\kern-.08em
    T\kern-.1667em\lower.7ex\hbox{E}\kern-.125emX}}
\begin{document}

\history{}
\doi{}

\title{Receiver Design for Faster-than-Nyquist Signaling: Deep-learning-based Architectures}
\author{\uppercase{Peiyang Song}\authorrefmark{1}, \IEEEmembership{Student Member, IEEE},
\uppercase{Fengkui Gong}\authorrefmark{1}, \IEEEmembership{Member, IEEE}, \uppercase{Qiang Li}\authorrefmark{1}, \IEEEmembership{Student Member, IEEE},
\uppercase{Guo Li} \authorrefmark{1},  \IEEEmembership{Student Member, IEEE},
\uppercase{Haiyang Ding}\authorrefmark{2},  \IEEEmembership{Member, IEEE}}
\address[1]{State Key Laboratory of Integrated Service Networks, Xidian University, Xi'an, China}
\address[2]{School of Information and Communications, National University of Defense Technology, Xi'an, China}
\tfootnote{This work was supported in part by the National Key Research and Development Program of China under Grant 2018YFE0100500, in part by the National Natural Science Foundation of Shaanxi Province under Grant 2019CGXNG-010, in part by the Youth Program of National Natural Science Foundation of China under Grant 61901325, in part by the China's Post-Doctoral under Grant 2018M640958, in part by the Fundamental Research Funds for the Central Universities under Grant 8002/20101196698, in part by the National Natural Science Foundation of Shaanxi Province under Grant 2019JQ-658, and in part by the Innovation Fund of Xidian University 5001-20109195456.}


\corresp{Corresponding author: Fengkui Gong (e-mail: fkgong@xidian.edu.cn).}

\begin{abstract}
    Faster-than-Nyquist (FTN) is a promising paradigm to improve bandwidth utilization at the expense of additional intersymbol interference (ISI). In this paper, we apply state-of-the-art deep learning (DL) technology into receiver design for FTN signaling and propose two DL-based new architectures. Firstly, we propose an FTN signal detection based on DL and connect it with the successive interference cancellation (SIC) to replace traditional detection algorithms.
    Simulation results show that this architecture can achieve near-optimal performance in both uncoded and coded scenarios. Additionally, we propose a DL-based joint signal detection and decoding for FTN signaling to replace the complete baseband part in traditional FTN receivers.
    The performance of this new architecture has also been illustrated by simulation results. Finally, both the proposed DL-based receiver architecture has the robustness to signal to noise ratio (SNR). In a nutshell, DL has been proved to be a powerful tool for the FTN receiver design.
\end{abstract}


\titlepgskip=-15pt

\maketitle

\section{Introduction}
\label{sec:introduction}
\IEEEPARstart{T}{he} last couple of decades have seen the exponential
growth of wireless devices and data traffic. Nowadays, spectral efficiency has become extremely
valuable. With increasingly demanding requirements for spectral resources,
a promising technology named FTN is rediscovered
and has attracted a lot of attention in both industrial and academic
communities \cite{mazo1975faster, liveris2003exploiting, anderson2009new, bedeer2017very, fan2018mlse, sugiura2013frequency, ganhao2014high, ishihara2017iterative, song2019blind, chang2015tomlinson, chang2015faster, chang2016robust, maso2016pre, li2019symbol, li2018reduced}.

As known, in conventional Nyquist-criterion transmission, when available bandwidth
is $W$ Hz, the symbol interval $T$ is always set as $T\ge T_{N}=1/(2W)$.
The strict orthogonality between transmitted symbols guarantees the
signal recovery in the receiver. In contrast, the symbol interval
reduces to $T<T_{N}$ in FTN signaling to achieve a higher transmission
rate, which, at the same time, destroys the orthogonality and introduces
unvoidable ISI. Although the additional interference increases the
complexity to recovery original signals in the receiver, the Mazo
limit \cite{mazo1975faster} proves that without the expansion of
bandwidth and loss of BER performance, the FTN signaling
can achieve an up to 25\% higher transmission rate than conventional Nyquist-criterion design in the additive white
Gaussian noise (AWGN) channel. 

\Figure[th](topskip=0pt, botskip=0pt, midskip=0pt)[scale=0.33]{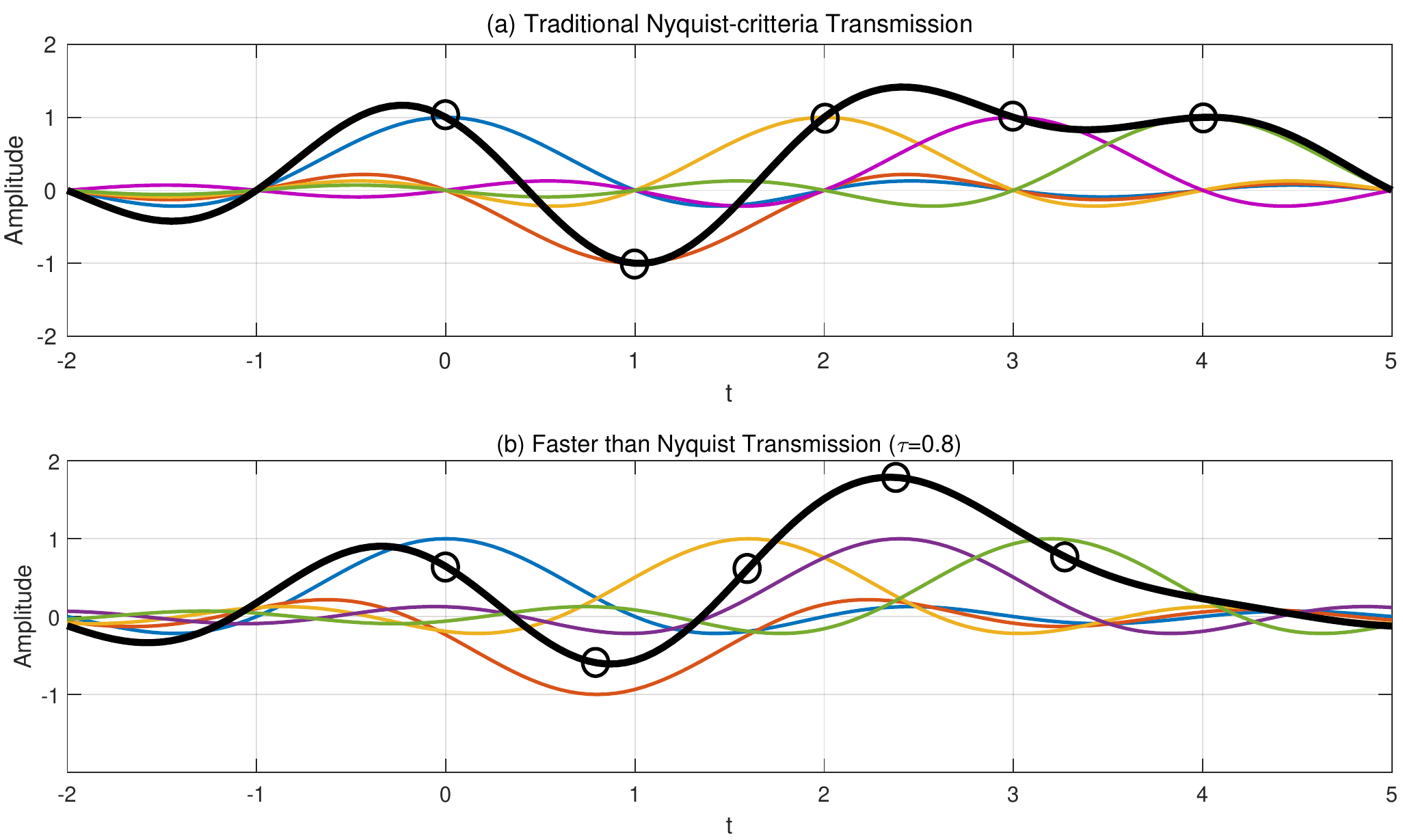}
{Waveform and the sampled symbols in Nyquist and FTN transmission. \label{fig:waveform}}

\Figure[tp](topskip=0pt, botskip=0pt, midskip=0pt)[scale=0.4]{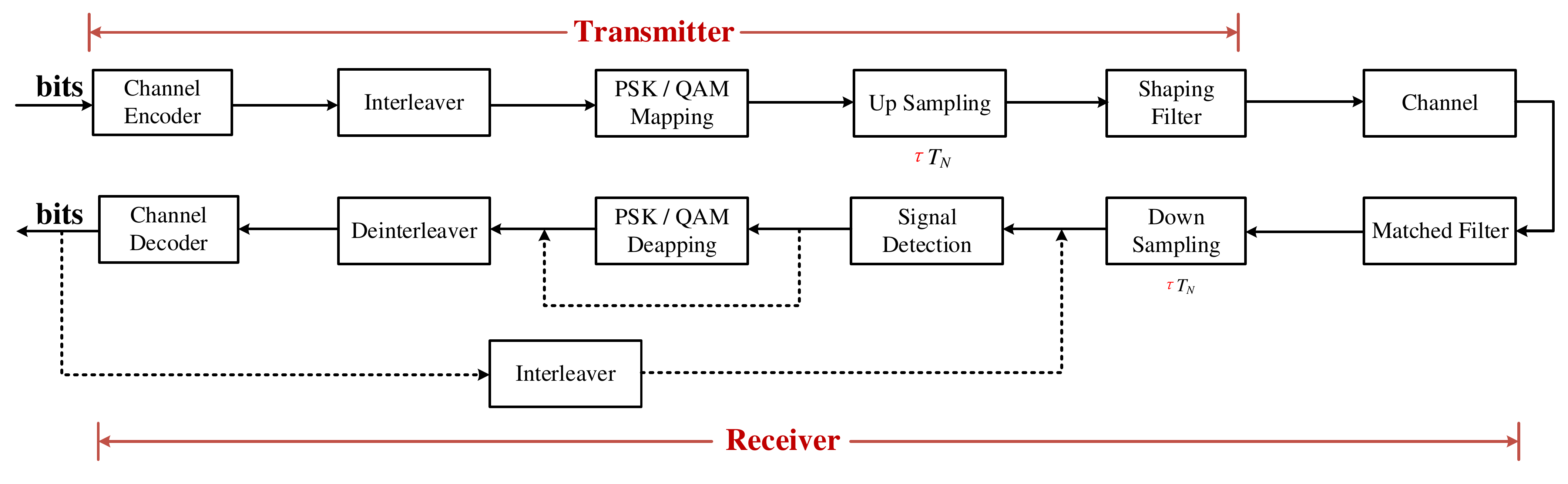}
{Block diagram of the traditional communication system with channel coded FTN signaling. \label{fig:conventional}}

Traditional receiver design focuses on the detection algorithms to eliminate the ISI caused by the smaller symbol interval. Among time-domain equalizations,  \cite{liveris2003exploiting} and its simplified version \cite{anderson2009new} formulate the FTN signal as convolutionally encoded symbols and applies the Viterbi algorithm for detection. \cite{bedeer2017very} employs a symbol-by-symbol signal detection, which achieves the near-optimal performance with very low complexity. Channel shortening is employed for maximum-likelihood sequence estimation (MLSE) in \cite{fan2018mlse}. 
Meanwhile, a few effective frequency-domain detection algorithms for FTN can also be found in the literature. \cite{sugiura2013frequency} designs a minimum mean square error (MMSE) frequency-domain equalization (FDE) for FTN detection. An H-ARQ technique for FTN signaling is proposed in \cite{ganhao2014high}. And \cite{ishihara2017iterative} focuses on iterative FDE architectures to eliminate the introduced ISI. Recently, an interesting work that focuses on the blind estimation for the packing ratio in FTN transmission has been addressed \cite{song2019blind} and is promising to make the receiver design more adaptive.



In recent years, a new trend has appeared to merge the two technologies of communications and DL \cite{simeone2018very, wang2017deep, he2019model}. Nowadays, DL has been widely employed in conventional communication scenarios, such as orthogonal frequency-division multiplexing
(OFDM) systems \cite{ye2018power, balevi2019one, gao2018comnet}, multi-antenna systems \cite{xue2018unsupervised}, channel estimation and prediction \cite{wang2018deep, yang2019deep, yang2019prediction}, channel coding \cite{nachmani2018deep, bennatan2018deep, cao2019deep}, modulation classification \cite{huang2019automatic}, etc. However, DL-based receiver design for FTN signaling, as far as we know, has not been studied yet in the literature.

In fact, signal detection for FTN based on sequence estimation (as well as the channel decoding) can be regarded as a classification problem which aims to divide a multiple dimension space into several parts. For example, when we try to recover $M$ transmitted symbols from $M$ received symbols in BPSK modulation, we practically divide an $M$-dimension space into $2^{M}$ parts. The power of DL in solving such classification problems has been proved by its successful application in image and voice recognition. Also, with the development of artificial intelligence (AI) chips \cite{basicmi2019}, the DL-based algorithms may show their advance in future communication systems. These facts have inspired us to employ DL into FTN receiver designs.


The contribution of this paper can be summarized as follows.
\begin{itemize}
    \item We propose joint DL-based detection and SIC to replace traditional FTN detection algorithms, where SIC is used to eliminate the interference and obtain more accurate log-likelihood ratio (LLR) values.

    \item We develop DL-based joint detection and decoding for FTN signaling, which can replace the whole baseband part of conventional FTN receivers with the DL-based architecture.

    \item We investigate the robustness of both the proposed receiver designs to SNR values. And results show that after training by the FTN data set under a specific SNR value, the proposed designs can fit the scenarios with different SNRs and achieve near-optimal performance in the offline recovery.

    \item We have carried out comprehensive evaluations to verify and analyze the proposed DL-based FTN receiver architectures.
\end{itemize}

Herein, we give the definition of notations which we will encounter
throughout the rest of the paper. Bold-face lower case letters (e.g.
$\boldsymbol{x}$) are applied to denote column vectors. Light-face
italic letters (e.g. $x$) denote scalers. $x_{i}$ is the $i$th
element of vector $\boldsymbol{x}$. $x(t) \ast y(t)$ denotes the convolution operation between $x(t)$ and $y(t)$. $\lfloor x\rfloor$ is the maximum integer less than or equal to $x$. And $P(\cdot)$ means the probability.

\Figure[tp](topskip=0pt, botskip=0pt, midskip=0pt)[scale=0.34]{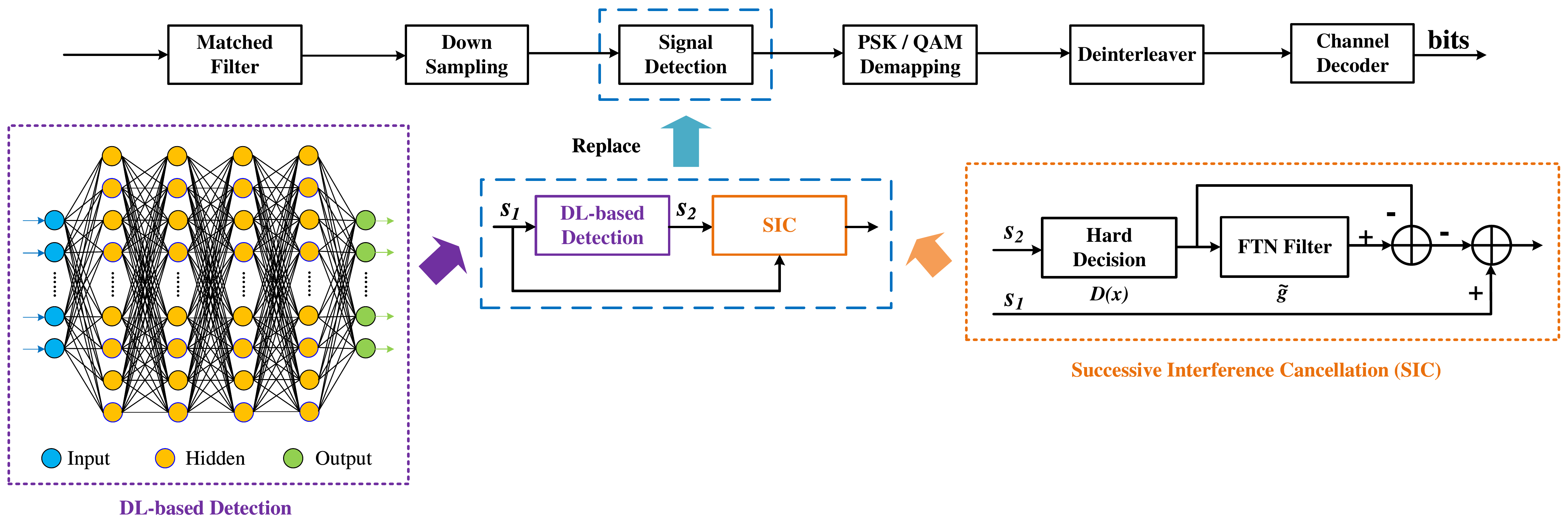}
{The architecture of the proposed FTN receiver with joint DL-based detection and SIC. \label{fig:DLDetection}}

\section{System Model}
We consider the communication system with the complex-valued quadrature amplitude modulation (QAM) scheme and AWGN
channel. As known, after constellation mapping, the baseband signal should
pass through a shaping filter $h(t)$. Hence, the transmitted signal
$s(t)$ can be written as

\begin{equation}
s(t)=\sqrt{E_{s}}\sum_{k=-\infty}^{+\infty}x_{k}h(t-k\tau T_{N}),
\end{equation}
where $E_{s}$ is the average energy of constellation symbols, $x_k$ $(k=0,\pm 1, \pm 2, \cdots)$ is the $k$th transmitted symbol and $\tau$ is the time acceleration factor which satisfies $\text{0 < \ensuremath{\tau\le1}}$.
Due to the existence of $\tau$, practical symbol interval $T$ is smaller
than Nyquist limit $T_{N}$, which helps the system achieve a higher
transmission rate.

Actually, when $\tau=1$, benefiting from the orthogonality between
any two symbols, each sample of the symbols will not be influenced
by the others. However, when $0<\tau<1$, each sample becomes a
weighted sum of different symbols, which will make it difficult to recover the transmitted signals. The intersymbol effect of FTN has been illustrated in Fig. \ref{fig:waveform}.

Fig. \ref{fig:conventional} shows the block diagram of the traditional communication system with channel coded FTN signaling, where $\tau T_N$ is not only the practical symbol interval of the symbols passing through the shaping filter but also the sampling interval for the signal which has passed through the matched filter. The LLR values can be calculated by PSK/QAM demapping as
\begin{equation} \label{eq:llr_equation}
    LLR_i=ln{\frac{P(x_i=0|\hat y_j)}{P(x_i=1|\hat y_j)}},
\end{equation}
where $i=\{1,2,3,...\}$, $j=\lfloor i/r \rfloor + 1$ and $r$ is the modulation order. $x$ and $\hat y$ represent the transmitted bits and detected symbols respectively.

The received signal after passing through the matched filter can be written as

\begin{align}
y(t) &= \left(s\left(t\right)+n\left(t\right)\right) \ast h(t)  \nonumber \\
 &=\sqrt{E_{s}}\sum_{k=-\infty}^{+\infty}x_{k}g(t-k\tau T_{N})+\widetilde{n}(t),
\end{align}
where $g(t)=\int h(x)h(t-x)dx$, $\widetilde{n}(t)=\int n(x)h(t-x)dx$,
and $n(t)$ is a zero mean complex-valued Gaussian random process
with variance $\sigma^{2}$. Throughout this letter, time synchronization
error is not taken into consideration. Hence, the $n$th sample of
received signal $y(t)$ can be obtained as (\ref{eq:received_samples}).

As seen, each sample of the received waveform contains not only the
expected symbol but also the weighted sum of both its previous and
upcoming symbols. A key problem for receivers is eliminating the ISI
from both directions and recover the transmitted sequence $\boldsymbol x$ from the received symbols $\boldsymbol y$. It will certainly increase the complexity of signal detection, which can be
regarded as the price of the higher transmission rate.

\begin{align} \label{eq:received_samples}
y_{n}  & =\sqrt{E_{s}}\sum_{k=-\infty}^{+\infty}x_{k}g(n\alpha T_{N}-k\alpha T_{N})+\widetilde{n}(n\alpha T_N)\nonumber \\
& =\underset{ISI\,from\,previous\,L-1\,symbols}{\underbrace{\sqrt{E_{s}}\sum_{k=-\infty}^{n-1}x_{k}g\left(\left(n-k\right)\alpha T_{N}\right)}}+\sqrt{E_{s}} x_{n}g(0)\nonumber \\
& \quad+\underset{ISI\,from\,upcoming\,L-1\,symbols}{\underbrace{\sqrt{E_{s}}\sum_{k=n+1}^{+\infty}x_{k}g\left(\left(n-k\right)\alpha T_{N}\right)}}+\widetilde{n}\text{\ensuremath{\left(n\alpha T_N)\right)}}.
\end{align}

\Figure[tp](topskip=0pt, botskip=0pt, midskip=0pt)[scale=0.3]{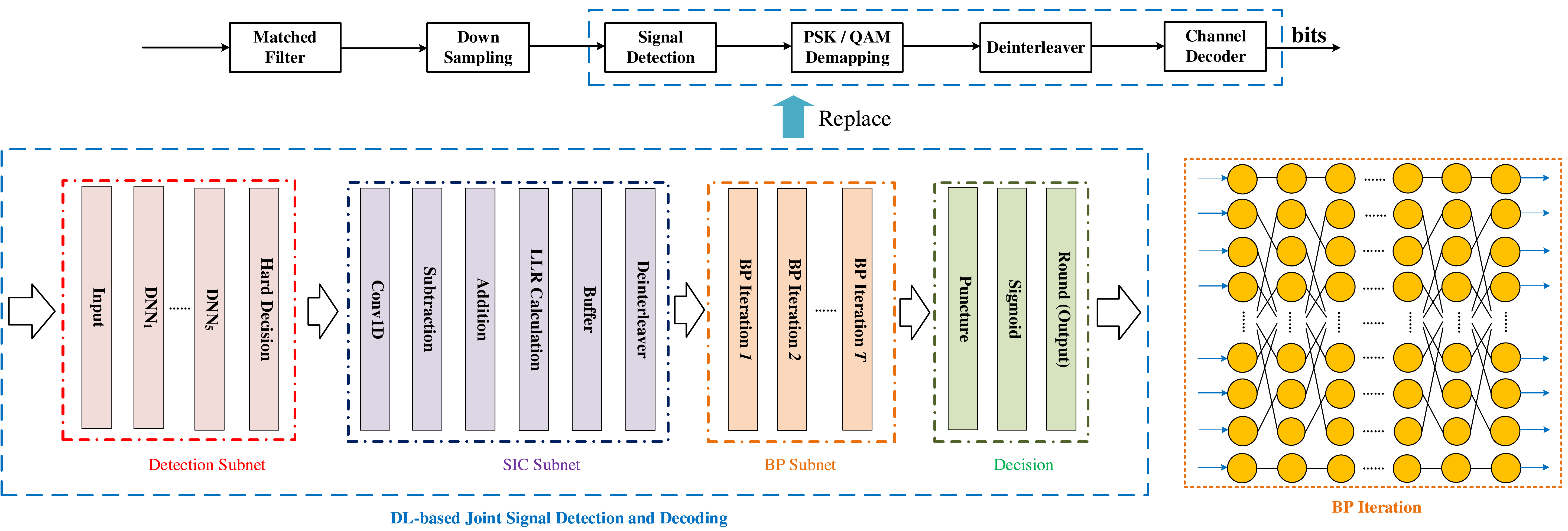}
{The architecture of the proposed FTN receiver with DL-based joint signal detection and decoding. \label{fig:DLJoint}}

\section{The Proposed FTN Receiver Design with DL-based Detection and SIC}\label{sec:DL-Based-FTN-Signal}
The proposed FTN receiver architecture with joint DL-based detection and SIC has been illustrated in Fig. \ref{fig:DLDetection}. The conventional signal detection is replaced by the proposed DL-based detection and SIC whose detailed structures have been shown in the part surrounded by the dotted line. In this section, we provide a detailed explanation of the proposed new architecture.

\subsection{DL-based Detection}
The proposed DL-based detection is essentially a deep neural network (DNN) which includes six layers (an input layer, an output layer and four hidden layers). As shown in Fig. \ref{fig:DLDetection}, each hidden layer is composed of a fully connected (FC) sublayer and a rectifier linear unit (ReLU) function while the output layer is simply an FC sublayer. The numbers of neurons in each layer are $L$, 320, 160, 80, 40 and $m$, respectively. $L$ and $m$ are sliding window parameters which will be further introduced in the next subsection. It should be noted that these numbers of layers and neurons are employed for all the scenarios in the following part of this paper and are applicable for other general FTN signalings. The performance gain by simply enlarging the size of the network is disappointing.

The input and output of the DL-based detection are real numbers. When BPSK is taken into consideration, one DNN should be employed on the real part of the received symbols. While in QPSK or higher QAM modulations, two identical structures (or multiplexing of one DNN) should be combined to detect both the real and imaginary parts of the received symbols.


\subsection{Dataset}

The training and testing data sets share the same structure. The input data set is composed of the real and imaginary parts of received symbols which have been downsampled and sliced by block length $L$ and step size $m$. And the label set contains the real and imaginary parts of corresponding transmitted symbols sliced by $m$. All these symbols are generated by the software simulation. Also, the structure of the data set has been shown visually in Fig. \ref{fig:SlidingWindow}.

\subsection{Workflow}

A notable feature of our proposed DL-based detection is the \emph{sliding
window}, as shown in Fig. \ref{fig:SlidingWindow}. The input in each
recovery is a sliding window with length $L$ on the real or imaginary part of the received symbols $\boldsymbol{y}$.
During each recovery, the DL-based detection tries to recover the
middle $m$ symbols of the input window. Then, the input window slides forward over $m$ symbols
to start the next recovery. This special architecture results from
the uncertainty of edge symbols of the input window. They always suffer from
severe ISI from either the previous or subsequent symbols while these symbols
are unknown since they are not contained in the input.

\Figure[th](topskip=0pt, botskip=0pt, midskip=0pt)[scale=0.33]{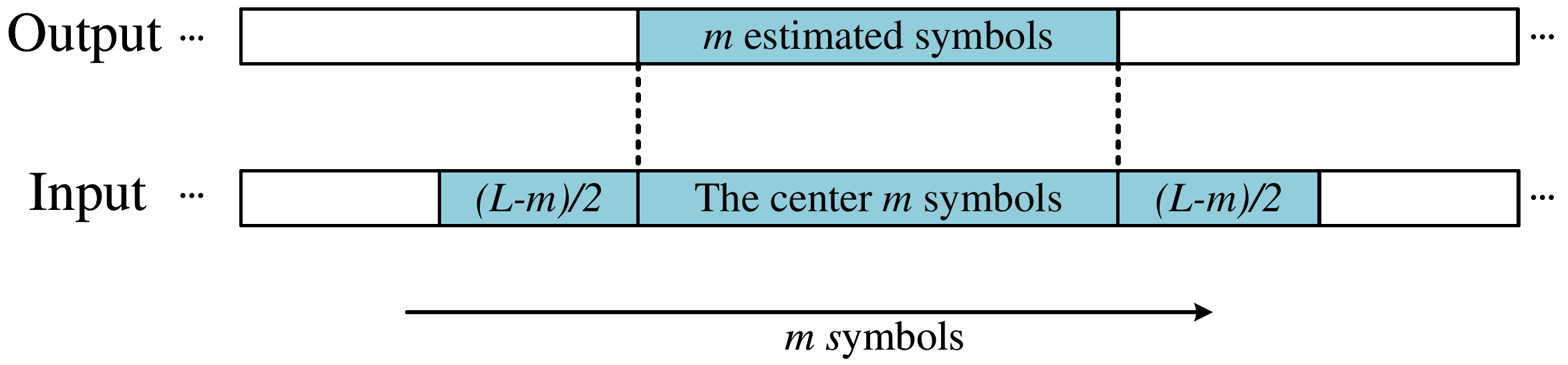}
{Workflow of the proposed DL-based signal detection for FTN signaling. \label{fig:SlidingWindow}}

Similar to most DL methods, our proposed detection
includes two stages named \textit{offline training} and \textit{online recovery}.
\begin{itemize}
\item \textit{Offline training:} During this stage, the model is trained
by known transmitted symbols and the corresponding received symbols which are generated with the given $\beta$ and $\tau$. The process called back-propagation is required to calculate the  loss function and its derivative to each neuron to update the network.
\item \textit{Online recovery}: During this stage, the DL-based detection can directly produce the estimated symbols by the received signals with the previously determined neural network. The back-propagation is not required, which can greatly reduce the delay and computational complexity of the detection.
\end{itemize}

It is worth noting that the two stages do not need to work iteratively. Once the offline training is completed, the neural network will be determined and does not need to be trained any more in the practical FTN detection. 

\subsection{Signal Reconstruction by SIC}

Benefiting from the correcting and anti-interference ability, channel coding, nowadays, has become a necessary part of wireless communications. To gain the benefit of large rate enhancement from the FTN signaling with start-of-the-art channel coding technologies, we introduce SIC to obtain more accurate LLR values from the proposed DL-based detection.
The main idea of the reconstruction is to calculate the interference and subtract it from the received symbols. The reconstructed symbols can be written as
\begin{equation} \label{eq:sic}
    \boldsymbol{\widetilde y} = \boldsymbol{s_1} -(D\boldsymbol{(s_2)}*\boldsymbol{\widetilde g}-D\boldsymbol{(s_2)}),
\end{equation}
where $\boldsymbol{s_1}$ and $\boldsymbol{s_2}$ represent the received symbols and detected symbols respectively, as illustrated in Fig. \ref{fig:DLDetection}. $D(\cdot)$ means the hard decision of a certain symbol sequence. The convolution operation is completed by the FTN filter, where the $\boldsymbol{\widetilde g}$ is the ISI vector and $\widetilde g_k=g\left((k-K)\tau T_N\right)$, $k=\{0, 1\cdots 2K\}$.

\section{The Proposed FTN Receiver Design with DL-based joint signal detection and decoding}\label{sec:DL-Based-Joint}
We propose a hybrid DL-based architecture, which has been illustrated in Fig. \ref{fig:DLJoint},  to replace both the signal detection and channel decoding in conventional FTN receivers. In this section, the proposed new architecture is detailed presented.

\subsection{DL-based joint signal detection and decoding}

In this section, we introduce the low-latency belief-propagation (BP) polar code decoder \cite{yuan2014early} into the FTN receiver design and proposed a DL-based complete baseband part in FTN receivers. The proposed DL-based joint signal detection and decoding is a cascade of a detection subnet, an SIC subnet, a BP subnet and finally a decision component. The detection subnet shares the same structure of the proposed DL-based detection in Section \ref{sec:DL-Based-FTN-Signal}. The SIC subnet  eliminates the interference among different symbols by (\ref{eq:sic}) with  a convolution layer, a subtraction layer and an addition layer. Besides, a buffer is employed to output a complete polar code block so that the BP subnet can work normally.

The BP subnet, which can be considered as the unfolding of the conventional BP decoding structure, contains several left-to-right (LTO) and right-to-left (RTL) propagations. The value of each layer can be obtained by

\begin{equation}
\left\{ \begin{array} { l } { L _ { i , j } ^ { t  } = g \left( L _ { i + 1,2 j - 1 } ^ { t  } , L _ { i + 1,2 j } ^ { t } + R _ { i , j + N / 2 } ^ {  t } \right) } \\ { L _ { i , j + N / 2 } ^ {  t  } = g \left( R _ { i , j } ^ {t} , L _ { i + 1,2 j - 1 } ^ {t} \right) + L _ { i + 1,2 j } ^ {t} } \\ { R _ { i + 1,2 j - 1 } ^ {t} = g \left( R _ { i , j } ^ {t} , L _ { i + 1,2 j } ^ {  t - 1 } + R _ { i , j + N / 2 } ^ {t} \right) } \\ { R _ { i + 1,2 j } ^ {t} = g \left( R _ { i , j } ^ {t} , L _ { i + 1,2 j - 1 } ^ {  t - 1 } \right) + R _ { i , j + N / 2 } ^ {t} } \end{array} \right.,
\end{equation}
where $L_{i,j}^t$ and $R_{i,j}^t$ represent the values of $j$-th node in $i$-th layer of RTL and LTR during the $t$-th iteration.  $g(x, y)=0.9375*\text{sign}(x) \text{sign}(y)\text{min}(\vert x\vert, \vert y \vert)$. Before running the iterations, the decoder should be initialized by
\begin{equation}
R _ { 1 , j } ^ { 1 } = \left\{ \begin{array} { l l } { 0 , } & { \text { if } j \in \mathcal { A } , } \\ { + \infty , } & { \text { if } j \in \mathcal { A } ^ { c }} \end{array} \right.
\end{equation}
and
\begin{equation}
L _ { n + 1 , j } ^ {1} =\frac { P \left( y _ { j } | x _ { j } = 1 \right) } { P \left( y _ { j } | x _ { j } = 0 \right) } ,
\end{equation}
where $\mathcal{A}$ and $\mathcal{A}^c$ are the information bits set and frozen bits set respectively.

Finally, a decision component is employed for the last layer to realize
\begin{equation}
f(x) = \left\{ \begin{array} { l l } { 0 , } & { x\le 0 } \\ { 1, } & { x > 0} \end{array} \right.,
\end{equation}
where the sigmoid function, which is used to limit the output to $[0, 1]$, can be written as
\begin{equation}
f(x) = \frac{1}{1+e^{-x}}.
\end{equation}

\subsection{Dataset}
 As can be seen from the previous analysis, only the detection subnet contains trainable parameters. Hence, it is not the whole model but simply the detection subnet that needs to be trained. And the dataset of training and testing has been provided in Section \ref{sec:DL-Based-FTN-Signal}.

\subsection{Workflow} \label{subsec:Workflow}
\Figure[th](topskip=0pt, botskip=0pt, midskip=0pt)[scale=0.33]{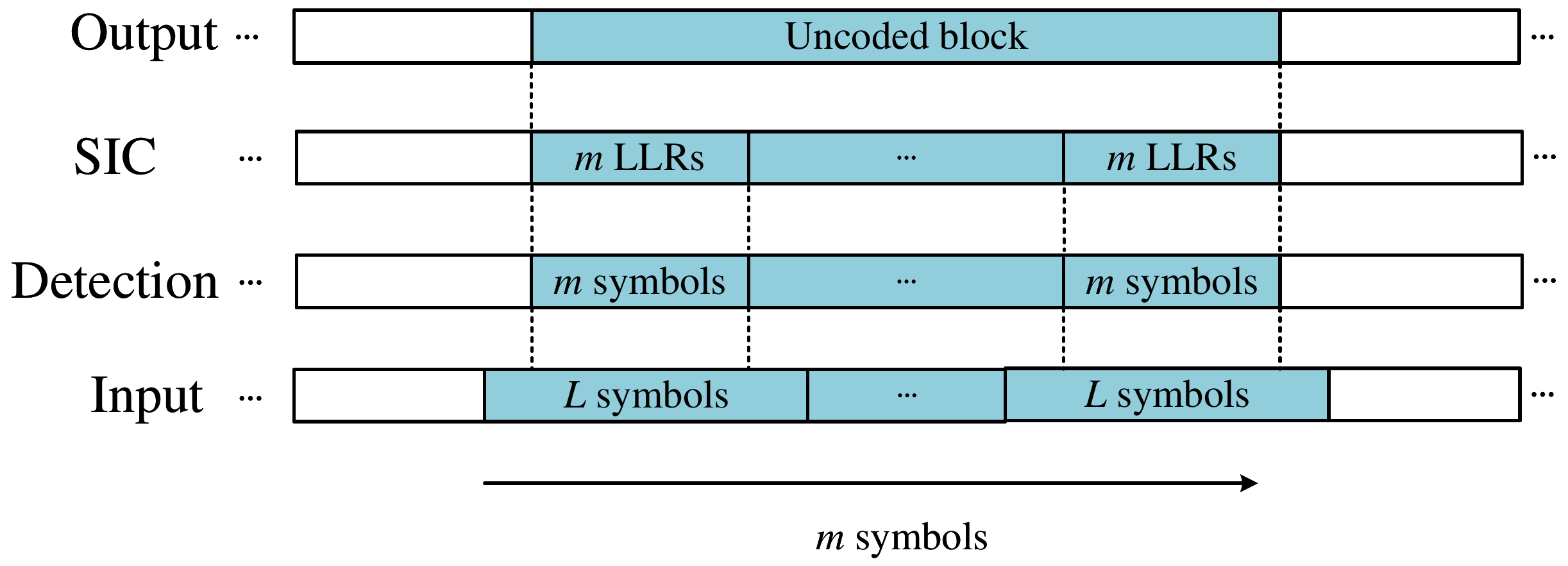}
{Workflow of the proposed FTN receiver design with the DL-based joint detection and decoding. \label{fig:codedslidingwindow}}
Fig. \ref{fig:codedslidingwindow} illustrates the workflow of the proposed DL-based joint signal detection and decoding. Similar to Section \ref{sec:DL-Based-FTN-Signal}, the input in each recovery is a sliding window with length $L$ on the received symbols while the output is the corresponding $m$ estimated symbols. Aided by the buffer in the SIC subnet, every $L_b/(R_br)$ LLRs are sent to the BP subnet, where $L_b$ is the code length and $R_b$ is the code rate. Finally, a complete uncoded bit block is obtained after BP and decision.

\section{Simulation Results}\label{sec:Simulation-Results}
In this section, we assess the performance and robustness of our proposed two DL-based FTN detection and decoding architectures. All the simulation results are obtained on the test data set. The square root raised cosine (SRRC) filters with different roll-off factors are taken into consideration. A more detailed list of the parameters is provided in Table \ref{table:parameters_DL_Detection}.

\begin{table}[th]
\newcommand{\tabincell}[2]{\begin{tabular}{@{}#1@{}}#2\end{tabular}}
    \renewcommand\arraystretch{1.5}
    \centering
    \caption{Training and testing parameters of the proposed DL-based FTN detection.}
    \label{table:parameters_DL_Detection}
    \setlength{\tabcolsep}{13pt}
    \begin{tabular}{|c|c|}
    \hline
    Parameter & Value \\
    \hline
    Training Data Size & $1.6\times10^9$ symbols \\\hline
    Training $E_b/N_0$ & $E_b/N_0$@\{BER=$2\times10^{-4}$\} \\\hline
    Learning rate & \tabincell{c}{0.001, 0.0002, 0.00004} \\\hline
    Loss function & Mean square error (MSE) \\\hline
    Optimizer & Adam \\\hline
    Testing Data Size & $3.2\times10^7$ symbols for each $E_b/N_0$ \\
    \hline
\end{tabular}
\end{table}


 \subsection{Robustness to SNR Mismatching}
 It is very important for the proposed DL-based detection to be robust to the SNR values, without which the proposed DL-based detection will be trained and employed for different SNR values independently and suffer from high complexity resulting from SNR estimation and the store of massive DL network parameters corresponding to different SNR values.

 Throughout a lot of simulations, fortunately, we find an interesting point that the DL network trained with $E_b/N_0@$ $\{BER=2\times10^{-4}\}$ (e.g. 7.9dB in QPSK) always shows the near-optimal performance in the offline prediction stage. $E_b/N_0@\{BER=x\}$ here means the $E_b/N_0$ value under which the ideal QAM modulation with Nyquist-criterion can achieve a BER performance of $x$ in the AWGN channel.

\subsection{Performance of the proposed DL-based detection in uncoded scenarios}
To better evaluate the performance of our proposed DL-based detection, we choose three traditional algorithms maximum a posteriori (MAP)\cite{li2018reduced}, MMSE FDE \cite{sugiura2013frequency} and SSSgbKSE \cite{bedeer2017very} as the baselines. And MAP algorithm is generally considered as the optimal detection scheme.

\Figure[th](topskip=0pt, botskip=0pt, midskip=0pt)[scale=0.57]{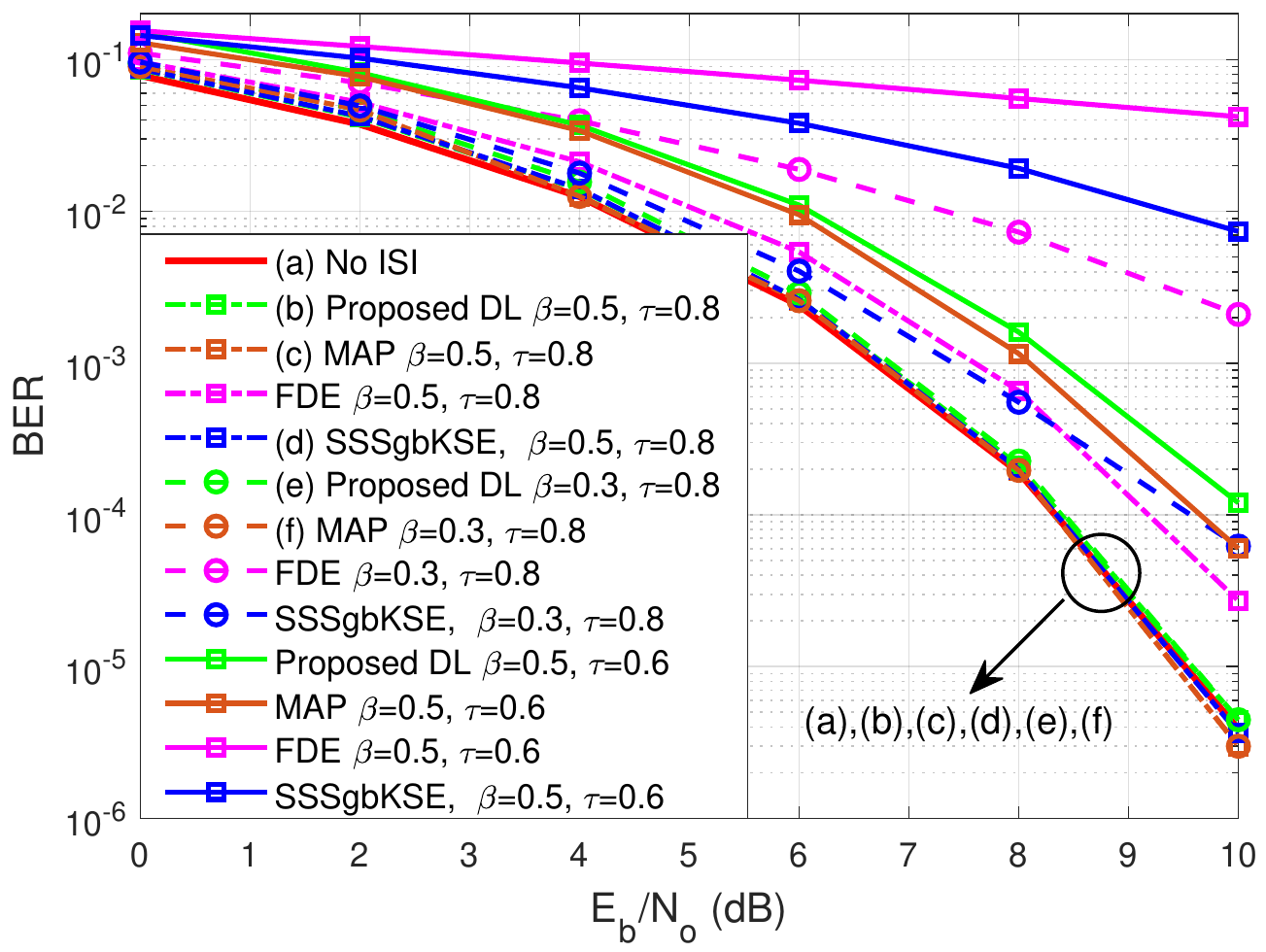}
{Performance of the proposed DL-based signal detection versus some baselines for uncoded FTN signaling.\label{fig:DLUncoded}}

Fig. \ref{fig:DLUncoded} compares the BER performance of our proposed DL-based detection with the baselines. As shown, all the detection schemes show similar performances when $\alpha=0.5$ and $\tau=0.8$. The SNR gain of our proposed DL algorithm over FDE is about 0.9dB, considering BER=$10^{-4}$. And as ISI becomes more severe, the SNR gains of the proposed DL algorithm over SSSgbKSE and FDE get larger. When $\alpha=0.5$ and $\tau=0.6$, the gains increase to more than 10dB, while the performance of the proposed DL algorithm and MAP just appears to differ and the distance of their SNR at BER=$10^{-4}$ is only 0.28dB. The simulation result confirms that the DL-based detection is applicable to FTN signal detection problems.

\subsection{Performance of the proposed DL-based Detection in High Order Modulations}

\Figure[th](topskip=0pt, botskip=0pt, midskip=0pt)[scale=0.55]{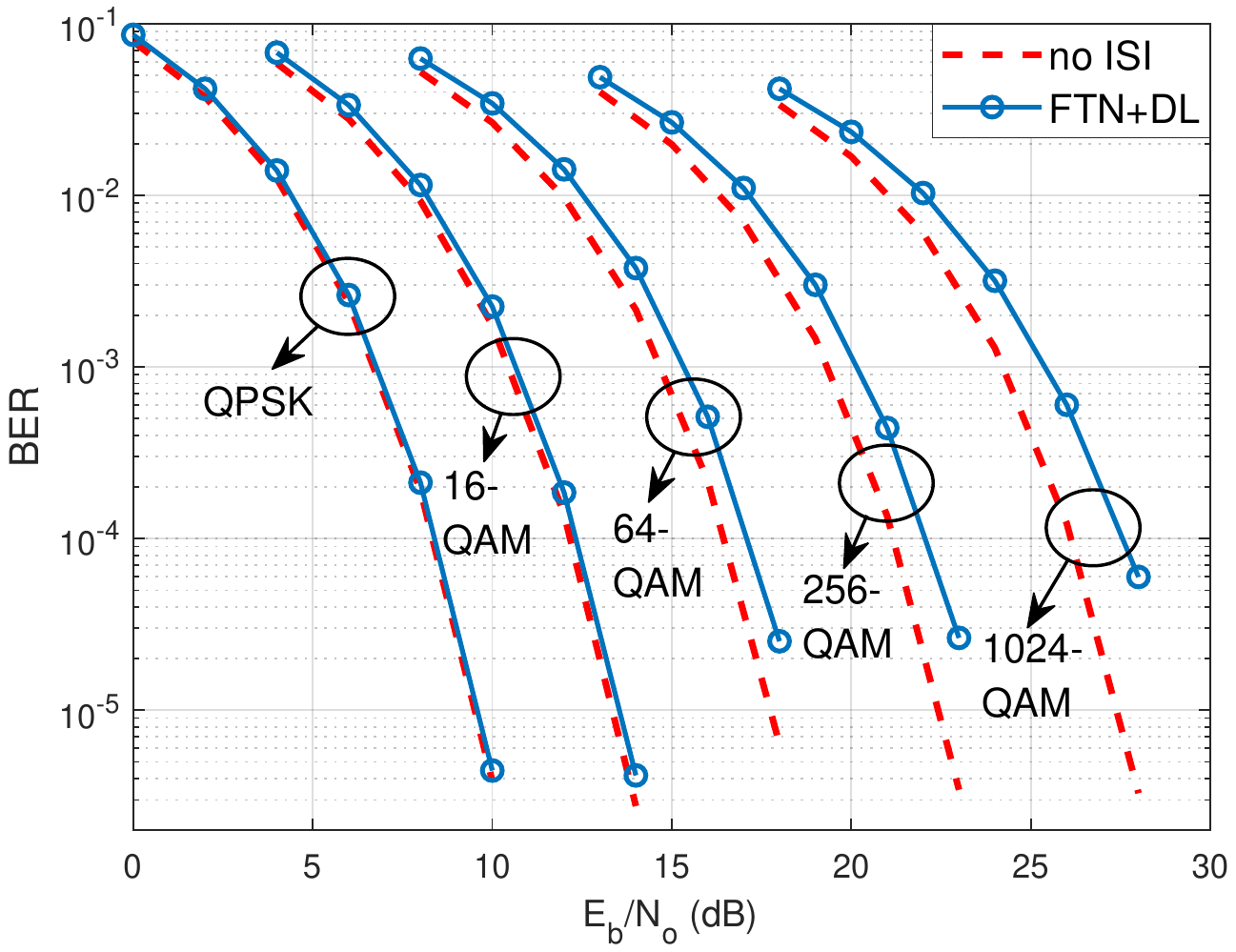}
{Performance of the proposed DL-based signal detection in high order modulations with $\beta=0.5$ and $\tau=0.8$. \label{fig:DLHighorder}}

Actually, signal detection for FTN will be difficult in high order modulations since the SNR gains are negligible. However, when parameters with relatively slight interference are employed, the FTN can still work well and contributes a rate enhancement. Fig. \ref{fig:DLHighorder} illustrates the BER performance of the proposed DL-based detection in high order modulations with $\beta=0.5$ and $\tau=0.8$, where the dashed lines represent the performance in conventional Nyquist-criterion systems. The result reveals the potential of DL-based detection for FTN signaling with high order modulations.

\subsection{Performance of the proposed receiver with joint DL-based detection and SIC}

\Figure[th](topskip=0pt, botskip=0pt, midskip=0pt)[scale=0.55]{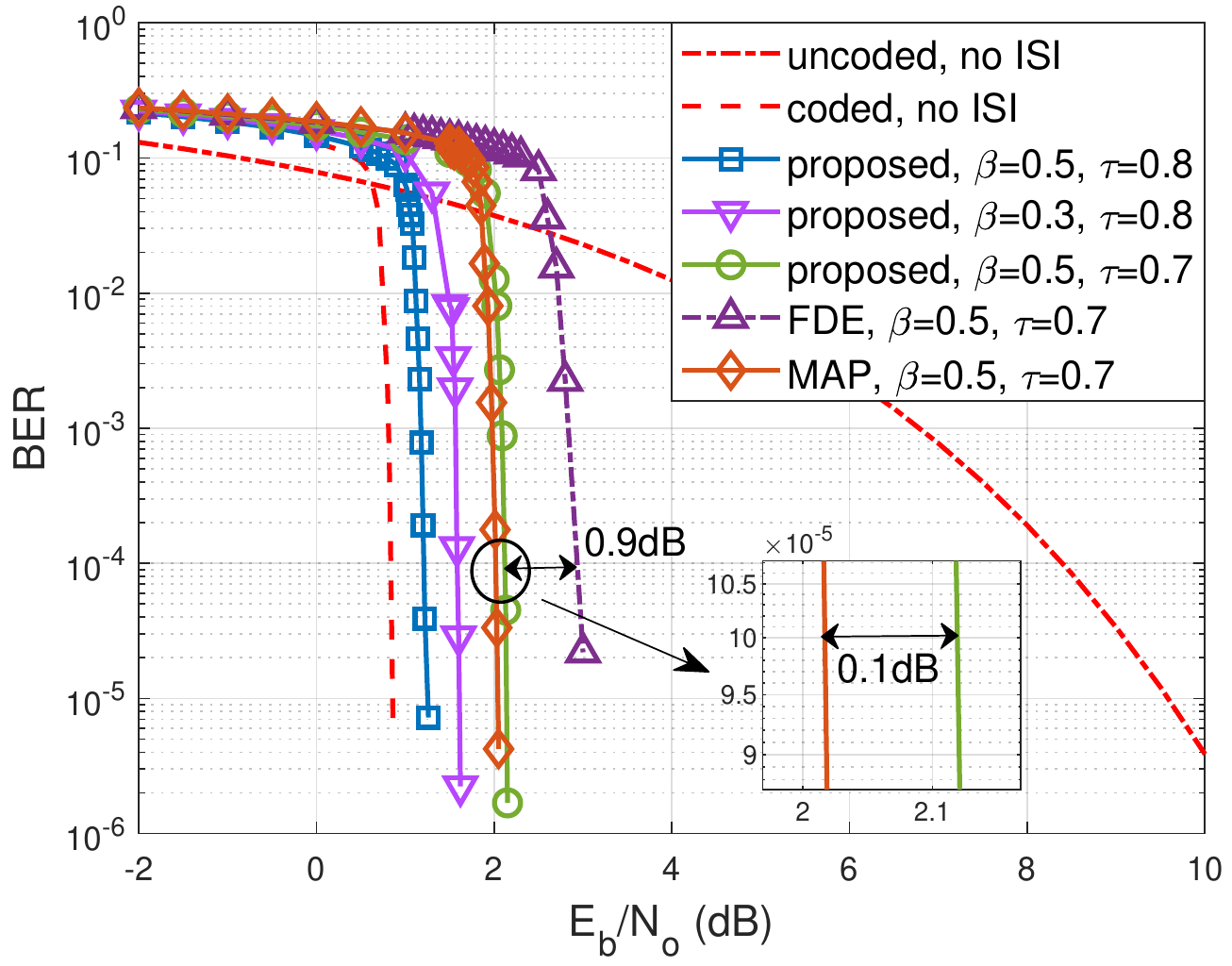}
{Performance of the proposed joint DL-based detection and SIC versus some baselines for (64800, 32400) LDPC-coded FTN signaling.\label{fig:DLLDPC}}

Fig. \ref{fig:DLLDPC} illustrates the performance of our proposed DL-based receiver design with detection and SIC versus other baselines in FTN signaling with (64800, 32400) LDPC code \cite{dvbs2xstandard}. The dashed lines show the performance in conventional coded and uncoded Nyquist-criterion systems. Since the SSSgbKSE method can not produce soft information independently (as stated in \cite{bedeer2017very}), it is not taken into consideration here.

As shown, the proposed receiver design can achieve great performance gains with the help of the start-of-the-art channel coding scheme. Compared to other baselines, the proposed design achieves an SNR gain over FDE by 0.9dB and is only 0.1 dB worse than MAP detection, considering $\beta=0.5$, $\tau=0.7$ and BER=$10^{-4}$. 
Moreover, the proposed DL-based detection and SIC can be directly cascaded with the existing channel decoders without changing their original structures or getting involved in their iterative decoding process, which is convenient for the practical implementation.

\subsection{Performance of the proposed receiver with DL-based joint detection and decoding}

\Figure[th](topskip=0pt, botskip=0pt, midskip=0pt)[scale=0.55]{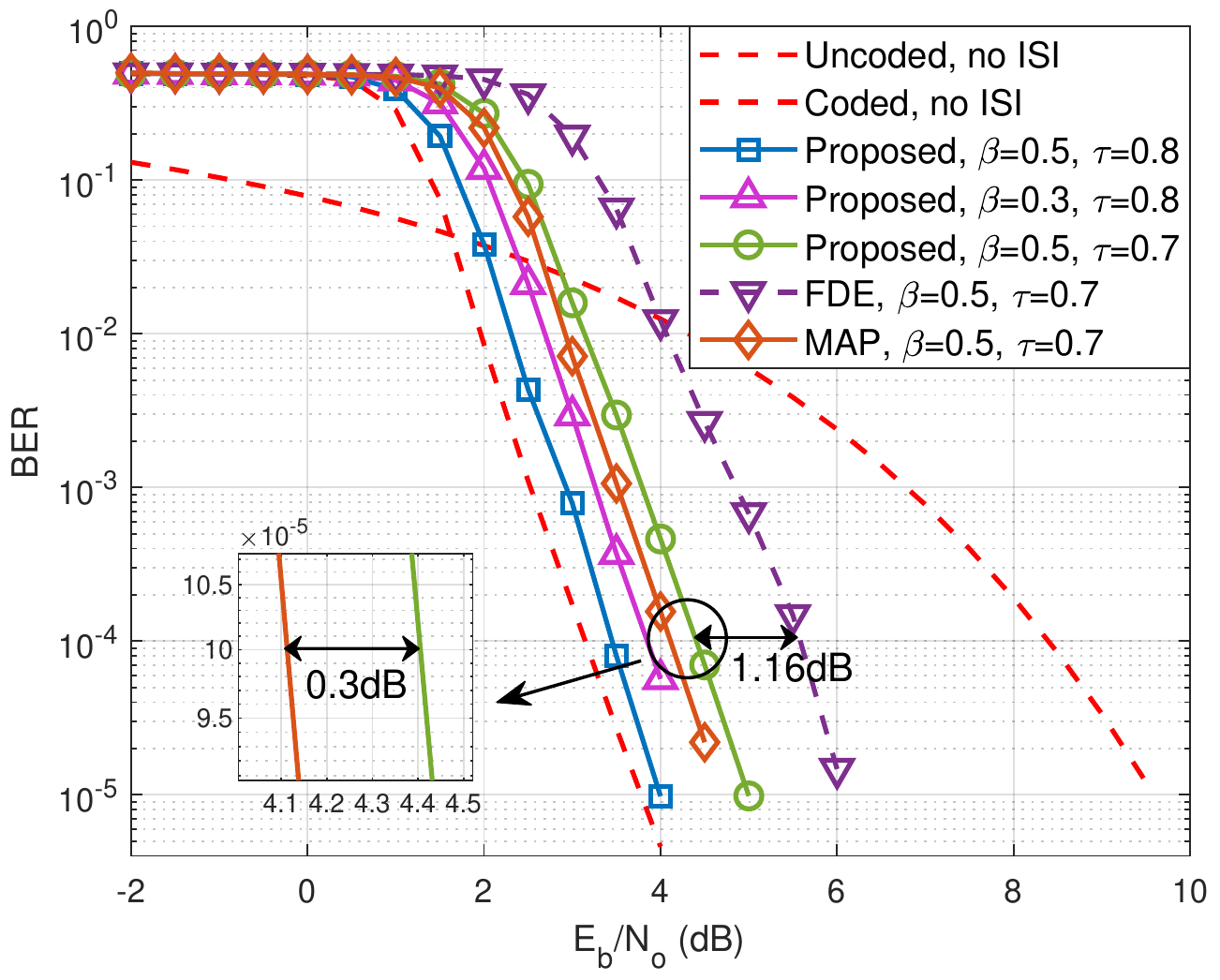}
{Performance of the proposed DL-based joint signal detection and decoding versus some baselines for (1024, 512) polar-coded FTN signaling. \label{fig:DLPolar}}

Fig. \ref{fig:DLPolar} illustrates the performance of our proposed DL-based joint detection and decoding with (1024, 512) polar code \cite{zhang2014practical}. The BP subnet with 50 iterations is taken into consideration. And similar to Fig. \ref{fig:DLLDPC}, the dashed lines are the performance of conventional coded and uncoded  Nyquist-criterion systems.

As shown, compared to other baselines, the proposed DL-based joint detection and decoding can achieve an SNR gain over FDE by 1.16dB and is only 0.3dB worse than MAP detection, considering $\beta=0.5$, $\tau=0.7$ and BER=$10^{-4}$. Additionally, the complete DL-based receiver design is applicable to the implementation by AI chips in the future.

\subsection{The complexity analysis of different FTN detections}
\begin{table}[th]
	\renewcommand\arraystretch{1.5}
	\centering
	\caption{Computational complexity of different FTN detection schemes.}
	\label{table:complexity_comparison}
	\setlength{\tabcolsep}{10pt}
	\begin{tabular}{|c|c|c|c|}
		\hline
		Scheme & Additions & Multiplications & Parallel \\\hline
		FDE\cite{sugiura2013frequency} & 8194 & 8196 & Support \\\hline
		Proposed DL & 9720 & 9720 & Support \\\hline
		MAP\cite{li2018reduced} & 12770 & 3980 & No \\\hline
	\end{tabular}
\end{table}

To compare the computational complexity of different FTN detections, for each scheme, we count the numbers of addition and multiplication operations consumed for every single estimated symbol and list them in Table \ref{table:complexity_comparison}. Also, the parallelizability of each scheme is taken into consideration to show whether the parallel implementation can be carried out to convert the computational complexity into space occupation to reduce the
detection delay. In view of the fact that the offline training does not involve in the FTN
signal detection in practical communication systems, as described in Section \ref{sec:DL-Based-FTN-Signal}, we do not include the training process into the evaluation about computational complexity of the proposed DL algorithm.

As shown, among the schemes, the complexity of the proposed DL algorithms is very close to that of FDE detection. And both these two schemes can be implemented in parallel to reduce the detection delay. Although MAP algorithm requires fewer multiplications than FDE and the proposed DL algorithm, the lack of support for parallel implementation will lead to severe detection delay, which has been proved by its application in turbo decoders.

\section{Conclusions}\label{sec:Conclusions}

FTN is a promising technology to improve spectrum efficiency. This paper, as far as we know, is the first attempt to apply DL into channel coded FTN detection and decoding. In this paper, we propose two DL-based FTN receiver designs, which have shown a near-optimal performance over some traditional algorithms. In particular, the proposed DL-based joint detection and decoding architecture can replace the whole baseband part of the FTN receiver, which can effectively improve the integration of the FTN receiver design. Moreover, with the rapid development of AI chips, the proposed DL-based FTN receiver designs are promised to show their advantages in the future implementation for AI-aided communications. Finally, in our future work, we will conduct more researches to extend the DL-based FTN detection to other practical channel environments (e.g. the multi-path fading channel).





\bibliographystyle{IEEEtran}
\bibliography{database}

\EOD

\end{document}